\documentclass{PoS}
\usepackage{epsfig}
\newcommand{\ave}[1]{\left\langle #1 \right\rangle}

\newcommand{\order}[1]{ \mathcal{O} \left( #1 \right) }

\newcommand{\eqcomma}{\phantom{AA},\phantom{AA}}
\ShortTitle{Quantum ideal hydrodynamics on the lattice}
\title{Quantum ideal hydrodynamics on the lattice}
\author{Tommy Burch \\ Institut f\"ur Theoretische Physik, University of 
  Regensburg, 93040 Regensburg, Germany \\
  E-mail: \email{tommy.burch@physik.uni-r.de}}
\author{\speaker{Giorgio Torrieri}\\ FIAS, J.W. Goethe-Universit\"at, Frankfurt, Germany \\
  Pupin Physics Laboratory, Columbia University, New York\\
  E-mail: \email{torrieri@phys.columbia.edu}}
\abstract{
After discussing the problem of defining the hydrodynamic limit from 
microscopic scales, we give an introduction to ideal hydrodynamics in the 
Lagrange picture, and show that it can be viewed as a field theory, which 
can be quantized using the usual Feynman sum-over-paths prescription. 
We then argue that this picture can be connected to the usually neglected 
thermal microscopic scale in the hydrodynamic expansion. 
After showing that this expansion is generally non-perturbative, we show 
how the lattice can be used to understand the impact quantum and thermal 
fluctuations can have on the fluid behavior.}
\FullConference{Lattice 2013, 
  31st international symposium on Lattice Field Theory.  July 29-August 3 2013, Mainz, Germany}
\begin{document}

\section{Introduction}

At first sight, ``quantizing hydrodynamics'' appears nonsensical:  Hydrodynamics is usually defined as a classical theory, because it is an infrared effective description of ``many'' microscopic degrees of freedom which have equilibrated.
Nevertheless, as shown in \cite{hydro0,hydro1,DerivExp}, ideal hydrodynamics {\em can} be rewritten in quantum form.   Can we learn something useful from this exercise?

A lot has been written, recently, about the hydrodynamic limit of ``infinitely strongly coupled'' theories, and weather a ``quantum lower limit'' of viscosity over entropy density ($\eta/s$) exists.   While famous results in both the Boltzmann limit \cite{daniel} and the strongly coupled planar limit of a conformal field theory \cite{4pi} suggest the existence of such a lower bound at $\eta/s \sim \order{10^{-2}}$, a physical argument for such a limit's existence beyond such idealized setups is still lacking.
 
As argued in \cite{mooresound} and \cite{gthydro}, such an argument might come from the possibility, at ``infinitely small mean free path'', for {\em thermal} fluctuations to excite hydrodynamic degrees of freedom:
When viscosity is so low that ``typical'' sound waves, of frequency $\sim T$ and amplitude comparable to a thermal fluctuation, $\Delta \rho/\ave{\rho} \sim C_V/T^3$ (where $\rho$ is the energy density, $C_V$ the heat capacity and $T$ the temperature), survive for a time much larger than the thermal scale, $\sim 1/T$, Kubo's formula needs to be renormalized to account for the energy-momentum carried by the sound waves. 

While in the planar limit \cite{4pi} this contribution is negligible as it is $\order{N_c^0}$, at any finite degeneracy it will alter both the ``viscosity'' and the ``entropy density'', and, because of turbulence, such perturbations are not guaranteed to stay small without a cutoff.    

In the absence of a sizable microscopic cutoff, quantum mechanics becomes relevant: a typical turbulent evolution in four dimensions involves a cascade from high amplitude low frequency perturbations to low amplitude high frequency ones \cite{cascade}.   This tacitly assumes that one can conserve energy by, simultaneously, decreasing the amplitude and increasing the frequency indefinitely.  Classically, this is indeed possible.
However, by Planck's law, a sound wave can not have a transverse energy $\sim \omega c_s^{-2} \Delta \rho$ smaller than its frequency $\omega$ in natural units. Quantum mechanics and energy conservation, therefore, cut the Kolmogorov cascade at frequencies $\sim T$ in natural units.   
Since this cutoff is usually provided by viscosity, the existence of a quantum limit on viscosity is plausible from hydrodynamic arguments alone.

More generally \cite{betz}, it is often neglected that hydrodynamics is an expansion not in one small parameter, but two:   The most explored one is the dissipative Knudsen number, $\sim l_{mfp}/R \sim \eta/(sTR)$, the mean free path ($ l_{mfp} \sim \eta/(sT) \sim $ the sound wave dissipation length) over ``system size'' $R$.   The other parameter, the ``distance between microscopic degrees of freedom'' $\sim 1/(gT)$, where $g$ is the microscopic degeneracy, has to be $\ll l_{mfp}$ for the BBGKY hierarchy to converge (in \cite{4pi}, $N_c \rightarrow \infty$ gives the same effect).     Thus, hydrodynamics requires, in terms of the microscopic degeneracy $g$, the entropy $s$, and the viscosity $\eta$, that
\begin{equation}
\label{scales}
 \underbrace{l_{micro}}_{s^{-1/3} \sim 1/g^{1/3}T} \ll \underbrace{l_{mfp}}_{\eta/(sT)} \ll \underbrace{l_{macro}}_{R \sim \partial_\mu \rho}
\end{equation}
Keeping the mean free path small but {\em comparable} to $1/(gT)$ is precisely the limit where microscopic thermal fluctuations can excite sound-waves.   
This is an explored limit, yet it is relevant for systems such as heavy ion collisions and ultracold atoms, where the number of degrees of freedom is clearly nowhere near infinite, and $(T^3 V)^{-1}$ is not far from $(sTR)/\eta$.   
\section{The theory}
3D Ideal hydrodynamics with no conserved charges\footnote{We assume, for simplicity there are no conserved charges, so all ``density'' is energy density (any ``particles'' are balanced by antiparticles'').  Dense systems can be described by an extension of the approach described here \cite{DerivExp} } can be rewritten \cite{hydro0,hydro1} in terms of three fields $\phi^{I=1,2,3}$, which physically correspond to the $x,y,z$ coordinates of the comoving frame w.r.t. the lab frame.
 The choice of $\phi^{I=1,2,3}$ is of course not unique, as a perfect fluid is homogeneous, and in its comoving frame, invariant under rotations and rescalings.  This ambiguity can be represented as a symmetry, restricting the Lagrangian to the form
\begin{equation}
\label{lfluid}
L=F(B) = T_0^4 F \left( \mathrm{det}\left( B_{IJ} \right)\right)
\eqcomma
  B_{IJ} =   \partial^\mu \phi_I \partial_\mu \phi_J  
\end{equation}
The function $F(B)$ is left arbitrary, as it corresponds to different equations of state for the fluid.  

Dimensional analysis makes it apparent that the $F(B)$ should be defined, in terms of an energy scale $T_0$. 
$T_0$, in this context, must be the ``temperature'' of the microscopic degrees of freedom.   Equivalently, $T_0^{-1}$ means the distance at which such microscopic degrees of freedom become relevant.  Note that this tells us {\em only} about the density (and equilibrium/quantum fluctuations of it), and is in general different from the mean free path of the interacting theory, which in the ideal hydrodynamic limit goes to zero.

Thus, if Eq. \ref{lfluid} is used to build a partition function, the effective ``Planck's constant'' becomes dimensionful;   It is natural to identify $T_0^{-1}$ with the $l_{micro}$ parameter in Eq. \ref{scales}, since as $T_0 \rightarrow \infty$ the classical picture of hydrodynamics should emerge.   Eq. \ref{lfluid}, however, makes it clear the expansion in $T_0$ might be strongly non-perturbative, as can be expected from a ``turbulent'' thermally fluctuating ideal fluid.   
It is straight-forward to show that the energy-momentum tensor corresponding to the Lagrangian in Eq. \ref{lfluid} is that of ideal hydrodynamics
\begin{equation}
\label{ideal}
T_{\mu \nu} = (p+\rho) u^\mu u^\nu -p g^{\mu \nu}
\end{equation}
and hence this is simply an unusual reparametrization of ideal hydrodynamics.
The energy density and pressure in this notation are 
\begin{equation}
\rho = -F(B) \label{eosrho} \eqcomma
p = F(B) -2 B \frac{dF}{dB} \label{eosp}
\end{equation}
hydrodynamic flow is defined in terms of enenrgy flow as
\begin{equation}
\label{flow}
u^\mu = \frac{1}{6 \sqrt{B}} \epsilon^{\mu \alpha \beta \gamma}\epsilon_{I J K} \partial_\alpha \phi^I  \partial_\beta \phi^J   \partial_\gamma \phi^K
\end{equation}
We can also show that $\partial_\mu (\sqrt{B} u^{\mu})=0$.  By inspection, without any conserved charges (those are examined in \cite{DerivExp}) we can identify
\begin{equation}
s = gT_0^3 \sqrt{B}
\end{equation}
with the microscopic entropy.   Using the Gibbs-Duhem relation, then, the temperature will be 
\begin{equation}
T= \frac{e+p}{s} = T_0\frac{\sqrt{B}(dF/dB)}{g}
\end{equation}
The non-perturbative nature of the theory is confirmed by examining the ``vortex'' degrees of freedom \cite{hydro1}:  In a hydrostatic background, vortices {\em do not propagate}, yet carry arbitrarily small amounts of energy and momentum.   Thus, an S-matrix cannot be defined, since quantum vortices can survive for an arbitrarily long time, and dominate the vacuum (in the $T_0\rightarrow \infty$ limit {\em all} such quantum fluctuations are suppressed).
Thus, at finite $T_0$, the quantum expectation value of $T_{\mu \nu}$ could very well be different from the classical one, due to the backreaction of sound waves and vortices. 
On the lattice, one can investigate this independently of any perturbative expansion, and without deforming the theory in the infrared (as was done in \cite{hydro0,gthydro}).

Conversely, a lattice calculation could show the theory {\em is} deformed in the IR in a way that breaks some, if not most, of the lagrangian's symmetries.
An example for this in 2d classical hydrodynamics is the famous regular patterns of vortices that form in 2d fluids \cite{coherent}.   If quantum fluctuations generate solutions like this, the low energy theory will contain effective terms which break most of the symmetries of Eq. \ref{lfluid}, but will be missing in the perturbative expansion.

The theory formulated via Eq. \ref{lfluid} can be put on the lattice in the usual way, via
\begin{equation}
\ln \mathcal{Z} = \int \mathcal{D}\phi_I \exp \left(i \int d^4 x L +J \phi_I
\right) \underbrace{\rightarrow}_{lattice+Wick} \int d \phi_I^i \exp \left[-
  (a T_0 )^4 \sum_i  F(\phi_i) + J \phi_I \right]
\end{equation}
care needs to be taken since this theory is non-renormalizeable and, once
$F(B)$ has been defined, has no free parameters.     The existence of a
well-defined continuous limit is therefore not guaranteed. 
The scaling of observables with $T_0$ as one approaches the limit of $T_0 \rightarrow
\infty$ can however still be explored.   As we will investigate non-trivial quantum
structures in configuration space, such as quantum-seeded vortices, the large lattice limit will be as important as the continuum limit.
\section{Lattice implementation}

In order to ensure that the fields on all lattice sites participate in the 
same updating procedure, we use one-sided finite differences for derivatives 
and average over the eight per hypercube. 
The derivatives $\partial_\mu\phi^I$ and all quantities derived from them 
(e.g., $B_{IJ}$, $u_\mu$, $T_{\mu\nu}$) are thereby envisioned as occupying 
the centers of the hypercubes.

Since the fields represent the comoving coordinates of the fluid, it is 
better to use ``shifted'' variables to avoid problems with the periodic 
boundaries (i.e., one subtracts the hydrostatic background): 
\begin{equation}
  \pi^I = \phi^I - x^I \;\; \to \;\; 
  \partial_\alpha\phi^I = \partial_\alpha\pi^I + 1 \delta_\alpha^I \; .
\end{equation}

We expect this theory to describe quite extended structures (e.g., vortices) 
which are easily created from the ``vacuum'' and we therefore use HMC 
updates in an attempt to learn something about their amenability to change 
in the Markov process. 
Thus, we require the variation of the action with respect to the local field 
values: 
\begin{eqnarray}
  \frac{\delta S}{\delta\phi^I(x)} = 
  \frac{\delta S}{\delta \sqrt{B}} \frac{\delta \sqrt{B}}{\delta(\partial_\alpha\phi^J)} 
  \frac{\delta(\partial_\alpha\phi^J)}{\delta\phi^I(x)} 
\end{eqnarray}
\[\ = \sum_{y,\mu,\nu,\sigma} \frac{dF}{d\sqrt{B}} \delta^{IJ} 
  \delta(y-x\pm\hat\mu/2\pm\hat\nu/2\pm\hat\sigma/2) 
  \left. \frac{\sqrt{B}}{8} B_{JK}^{-1} \, |\epsilon_{\mu\nu\sigma\alpha}| \, 
  \partial_\alpha\phi^K \right|^{y - \hat\alpha/2}_{y + \hat\alpha/2} \; . 
\]

\begin{table}
  \label{RunTab}
  \begin{center}
    \begin{tabular}{|c|c|c|c|c|}
      \hline
      $L^4$ & $C (=aT_0)$ & traj & $d\tau_{MD}^{}$ & accept \\
      \hline
      $20^4$ & 0.8 & 4000 & 0.001 / 0.0005 & 49\% / 85\% \\
      $16^4$ & 1 & 10000 & 0.001 & 52\% \\
      $12^4$ & 1.33333 & 10000 & 0.0005 & 61\% \\
      $10^4$ & 1.6 & 10000 & 0.0005 & 41\% \\
      $8^4$ & 2 & 10000 & 0.00025 & 72\% \\
      $6^4$ & 2.66667 & 10000 & 0.00025 & 56\% \\
      \hline
    \end{tabular}
  \end{center}
\caption{\label{table1} Lattice parameters for runs with constant volume.}
\end{table}

We have implemented the updating algorithm and the calculation 
of observables in C code, with parallelization via OpenMP. 
Table \ref{table1}  displays a subset of our runs performed with 
the aim of preserving the physical volume: $L^4 = (16/(aT_0))^4$. 
To achieve reasonable acceptance rates, quite small values of 
the molecular-dynamics time step ($d\tau_{MD}$) are required 
(even at such small volumes). 
Correspondingly, convergence to a plateau (say, in 
$\langle \sqrt{B} \rangle$) is slow, typically requiring the omission 
of the first few thousand trajectories from observables and 
blocking of data to avoid autocorrelations. 
In future runs at larger volumes and lattice spacings 
(ideally, $L^{-1} \ll l_{macro}^{-1} \ll a^{-1} \ll T_0$), we may 
need to add mass terms to the scalar fields (or perhaps work at 
finite density: see Ref.~\cite{DerivExp}) and extrapolate to 
the massless (zero chemical potential) limit.

\section{Some preliminary results}

\begin{figure}[h]
\begin{center}
\includegraphics*[clip,width=4.5cm]{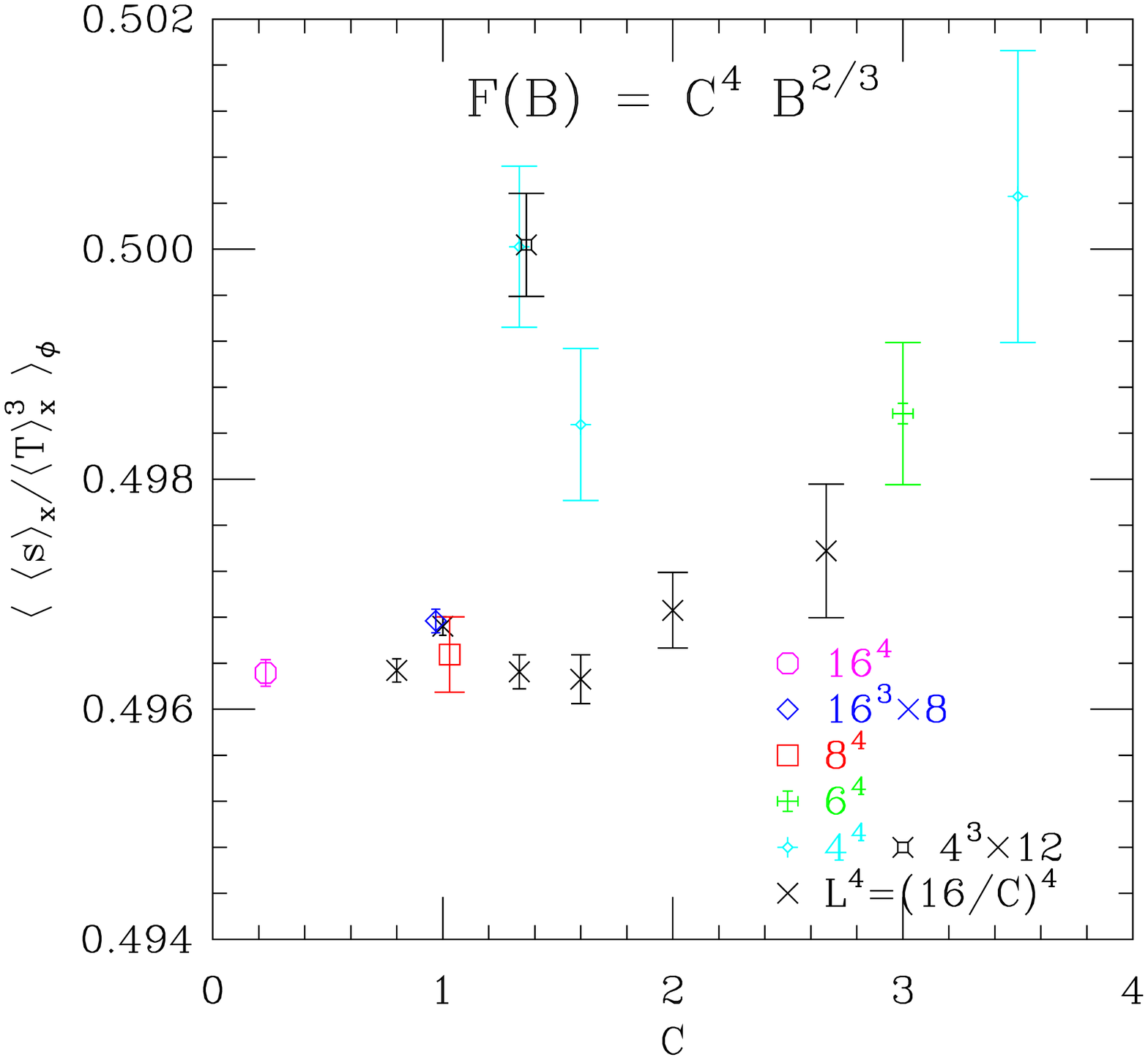}
\hspace*{0.5cm}
\includegraphics*[clip,width=4.5cm]{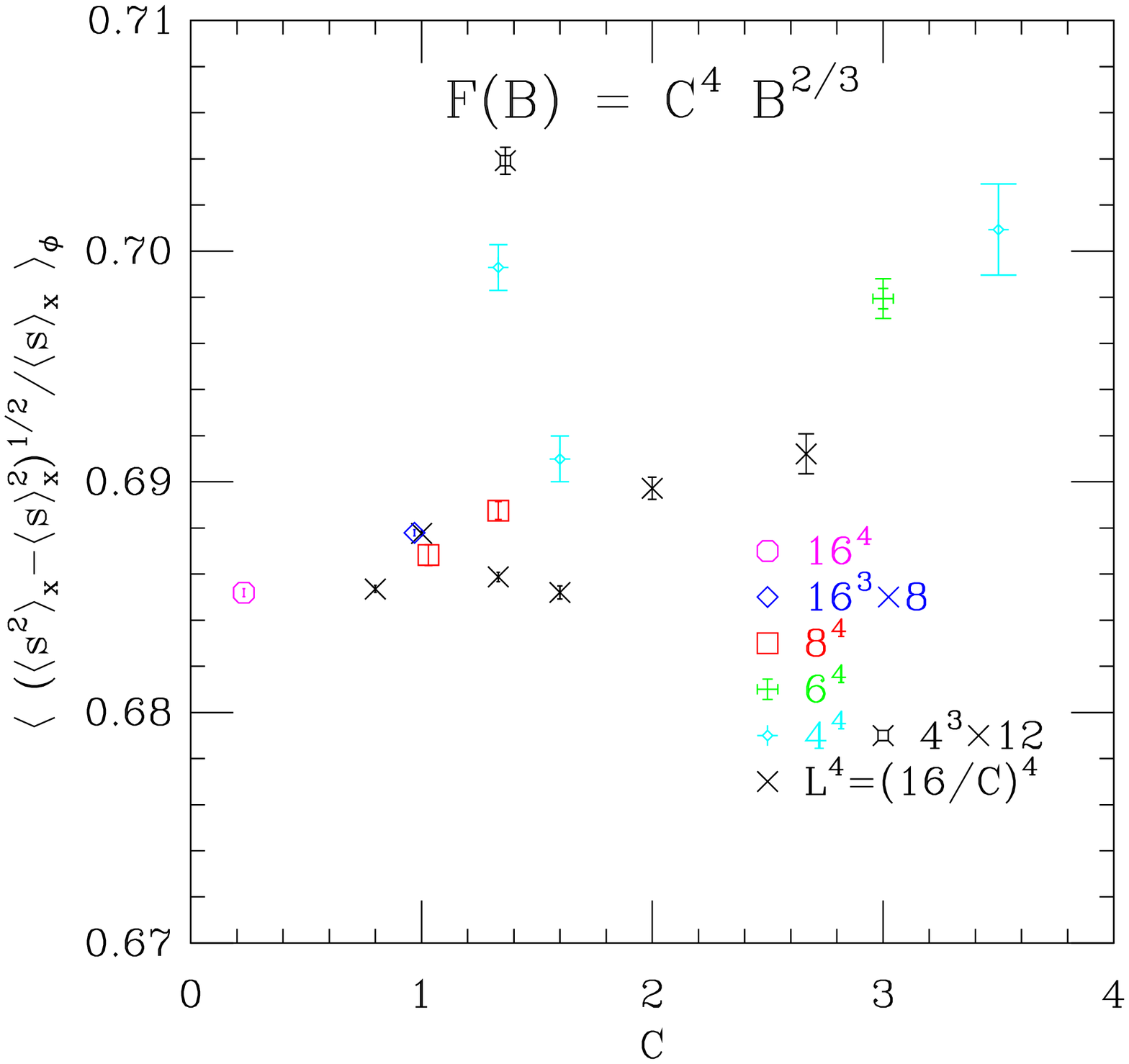}
\hspace*{0.5cm}
\includegraphics*[clip,width=4.5cm]{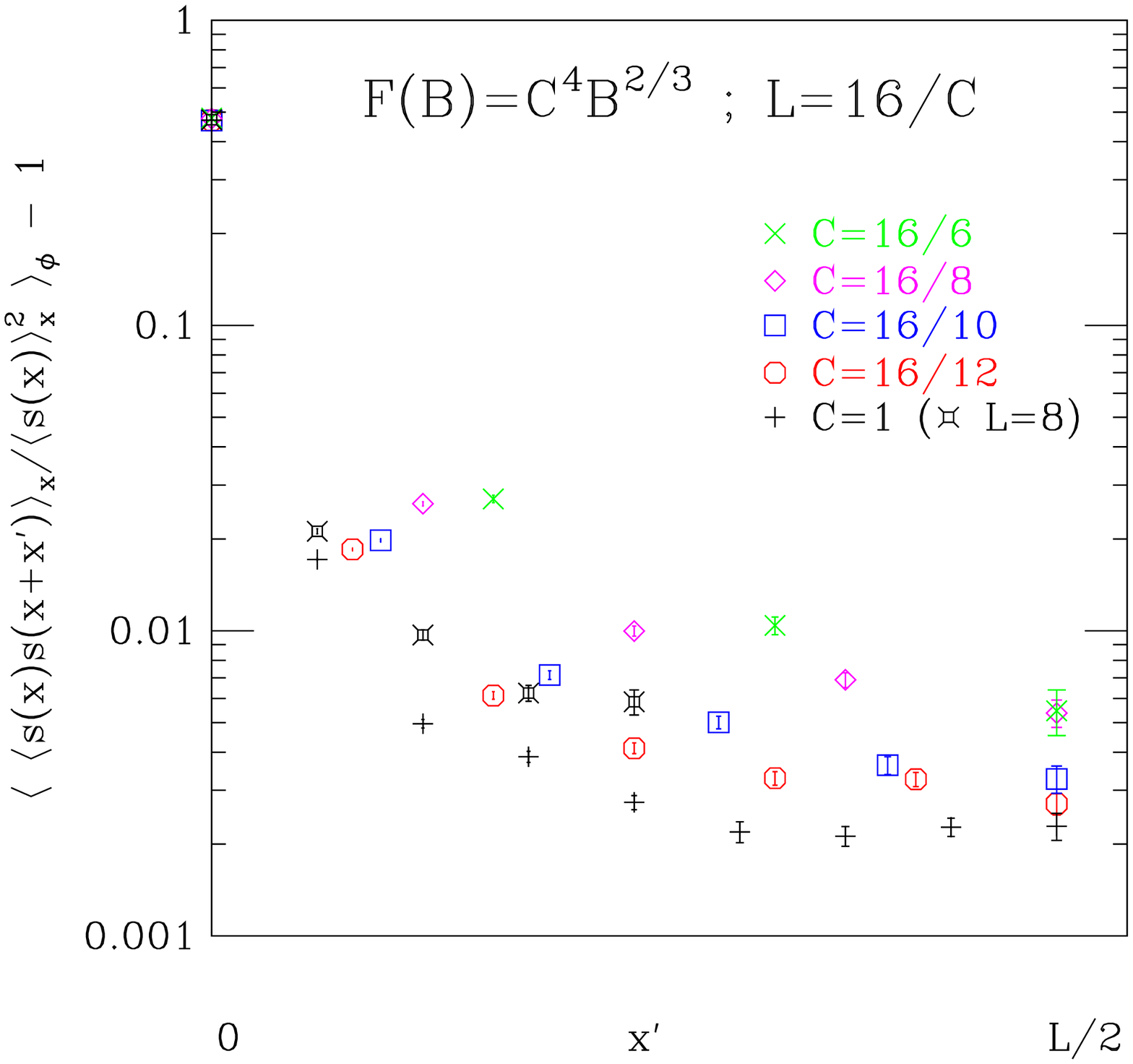}
\caption{\label{figs} The expectation value of the entropy density (left), 
the corresponding fluctuation (middle) 
and the relative spatial correlation along $\hat x$ (right).}
\end{center}
\end{figure}

A good initial example is the ideal gas EOS, 
\begin{equation}
F(B) = T_0^4 B^{2/3} \; . 
\end{equation}
This can be easily generalized to any monotonic EoS  (without phase transitions), for example an EoS ,say, fitting the QCD cross-over \cite{gthydro}.
The ideal gas, however, is a good testing laboratory as all of its parameters are very simple
\begin{equation}
\label{expect}
 \ave{e}= T_0^4 B^{2/3} = \frac{g\pi^2}{60} T^4 \eqcomma \ave{p}=\frac{\ave{e}}{3}  \eqcomma
 \ave{s}= T_0^3 \sqrt{B} = \frac{g\pi^2}{45} T^3  \eqcomma
 T =  \frac{4}{3g} T_0 B^{1/6}   
\end{equation}
This allows us to fill in all members of $T_{\mu \nu}$ in the static frame, $\ave{T_{\mu \nu}} = \mathrm{diag}\left[\ave{e},\ave{e}/3,\ave{e}/3,\ave{e}/3  \right]$
Correlations should be localized around standard thermodynamic fluctuations
\begin{equation}
\label{expectfluct}
\ave{e(x')e(x)}- \ave{e(x)}\ave{e(x')} = \delta^3\left( \vec{x}-\vec{x}' \right) C_V T^2 \sim \frac{4 \pi^2}{15} g T_0^8 
\end{equation}
Fig. \ref{figs} shows average and fluctuations of the $s$ observable. 
We have checked that the other observables scale with entropy as expected 
from Eq.~\ref{expect}.

%%%%%%%%%%%%%%%%%%%%%%%%%%%%%%%
%%%%%%%%%%%%%%%%%%%%%%%%%%%%%

The left panel of Fig.~\ref{figs} shows the entropy density as a function 
of the lattice spacing. 
Allowing for finite-size effects, no appreciable dependence is seen. 
The same goes for the rather sizable, relative fluctuations in the entropy 
density (middle plot). 
There also appears to be little dependence upon the ``macroscopic 
temperature'' $(aN_t)^{-1}$, i.e., the temperature of the phonon / 
vortex gas. 
It will be especially interesting to see if this trend (or lack thereof) 
continues as we push to larger lattice extents at larger $aT_0$ values: 
i.e., as we push from the microscopic-dominated entropy region to one 
where macroscopic structures carry a large share of entropy. 
The fluctuations do indeed lead to a ``correction'' to the input equation 
of state (EoS): the EoS dictates $s/T^3 = 27/64 \sim 0.422$, whereas 
averaging first over the lattice leads to 
$\ave{s}_x/\ave{T}_x^3 \sim 0.496$. 
One can see in the rightmost plot that the spatial correlations in 
the entropy density persist across the lattice (the same can be seen for 
$T$, $p$, and $\rho$). 
We cannot comment more on this result as yet, beyond remarking that it 
might show that quantum fluctuations are non-negligible for observables 
at {\em all} scales.

The ideal hydrostatic background further requires any space component 
of a vector or tensors observable's average should be zero. 
These can be parametrized into the flow tensor 
\begin{equation}
  \label{omega}
  \ave{\Omega_{\mu \nu}} =\ave{ u_\mu u_\nu + g_{\mu \nu}} =\ave{ B_{IJ}^{-1} 
    \partial_\mu \phi^{I}  \partial_{\nu} \phi^J} \; , 
\end{equation}
which gives insight to the turbulence seeded by quantum fluctuations. 
While, unlike the locally-defined scalar perturbation, the average 
$\Omega_{\mu \nu}$ is set by symmetries, its correlation function 
%$\ave{\Omega_{\mu \nu}(x) \Omega_{\mu \nu}(x')}$ 
can reveal interesting structures. 
In particular, a non-vanishing value at large distances can 
signal the vacuum generally contains ``quantum turbulence'', and can 
break some of the symmetries of the original Lagrangian.

%%%%%%%%%%%%%%%%%%%%%%%%%%%%%%%
\begin{figure}[h]
\begin{center}
\includegraphics*[clip,width=4.2cm]{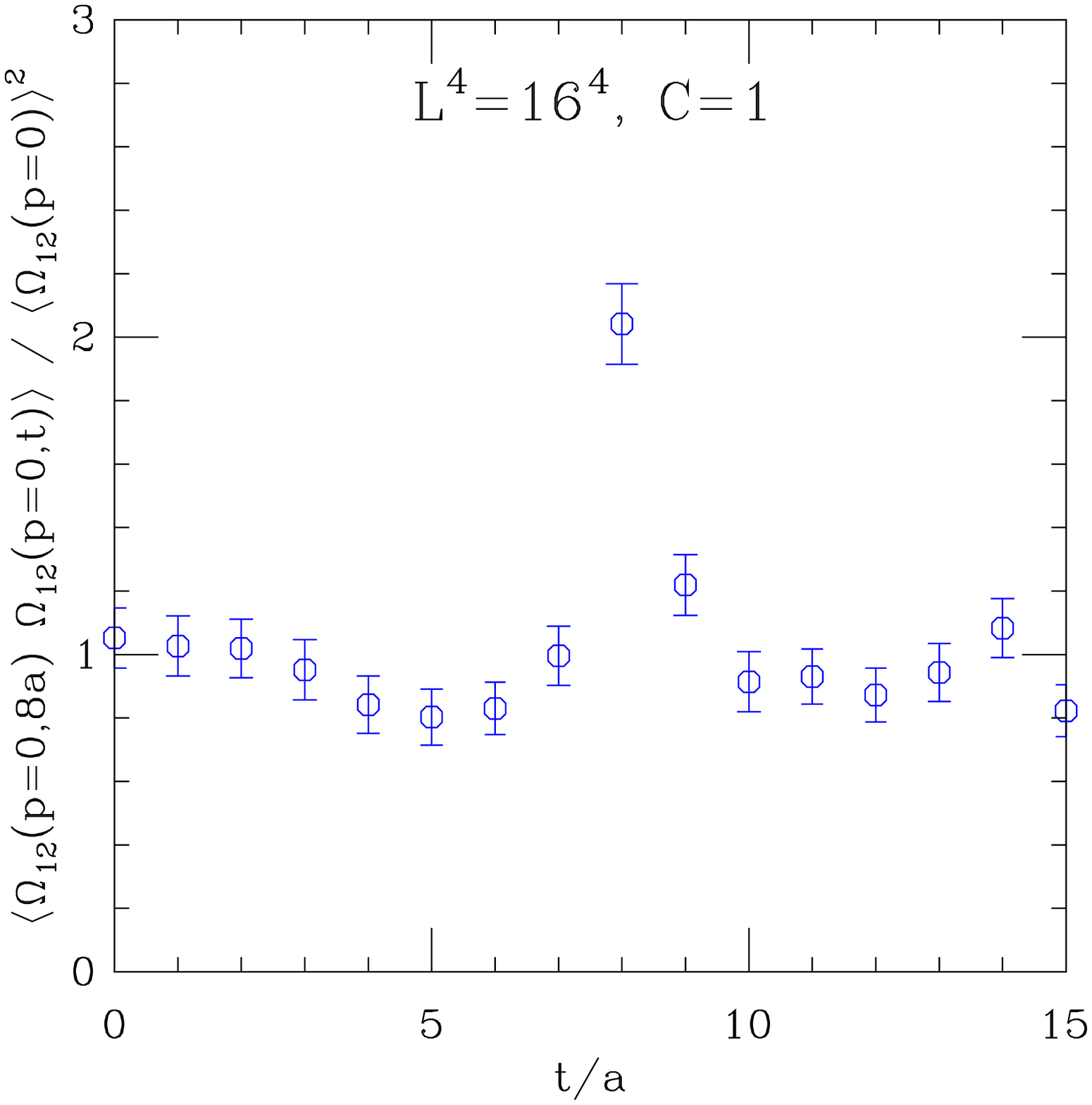}
\hspace*{0.5cm}
\includegraphics*[clip,width=4.5cm]{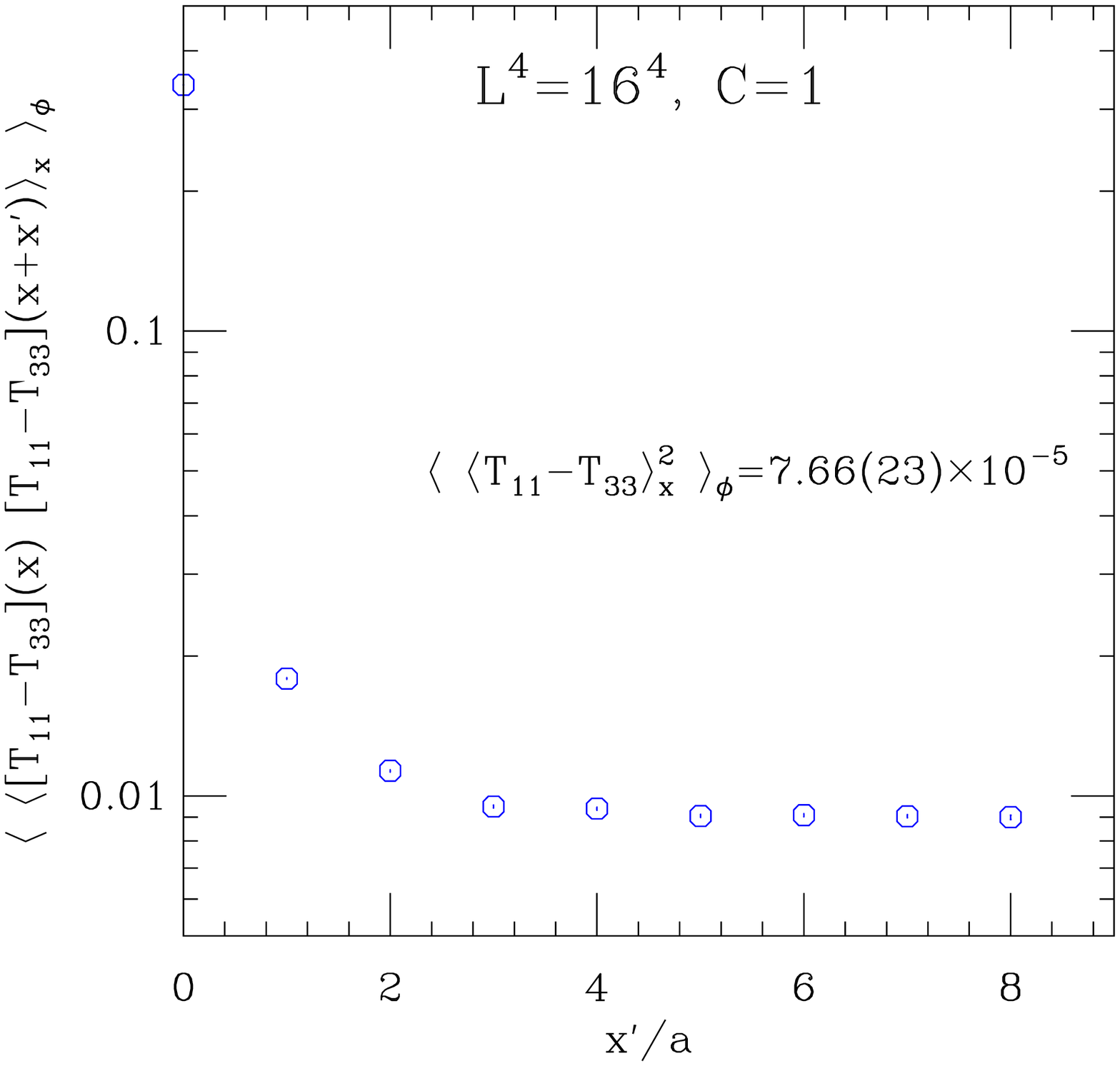}
\hspace*{0.5cm}
\includegraphics*[clip,width=4.6cm]{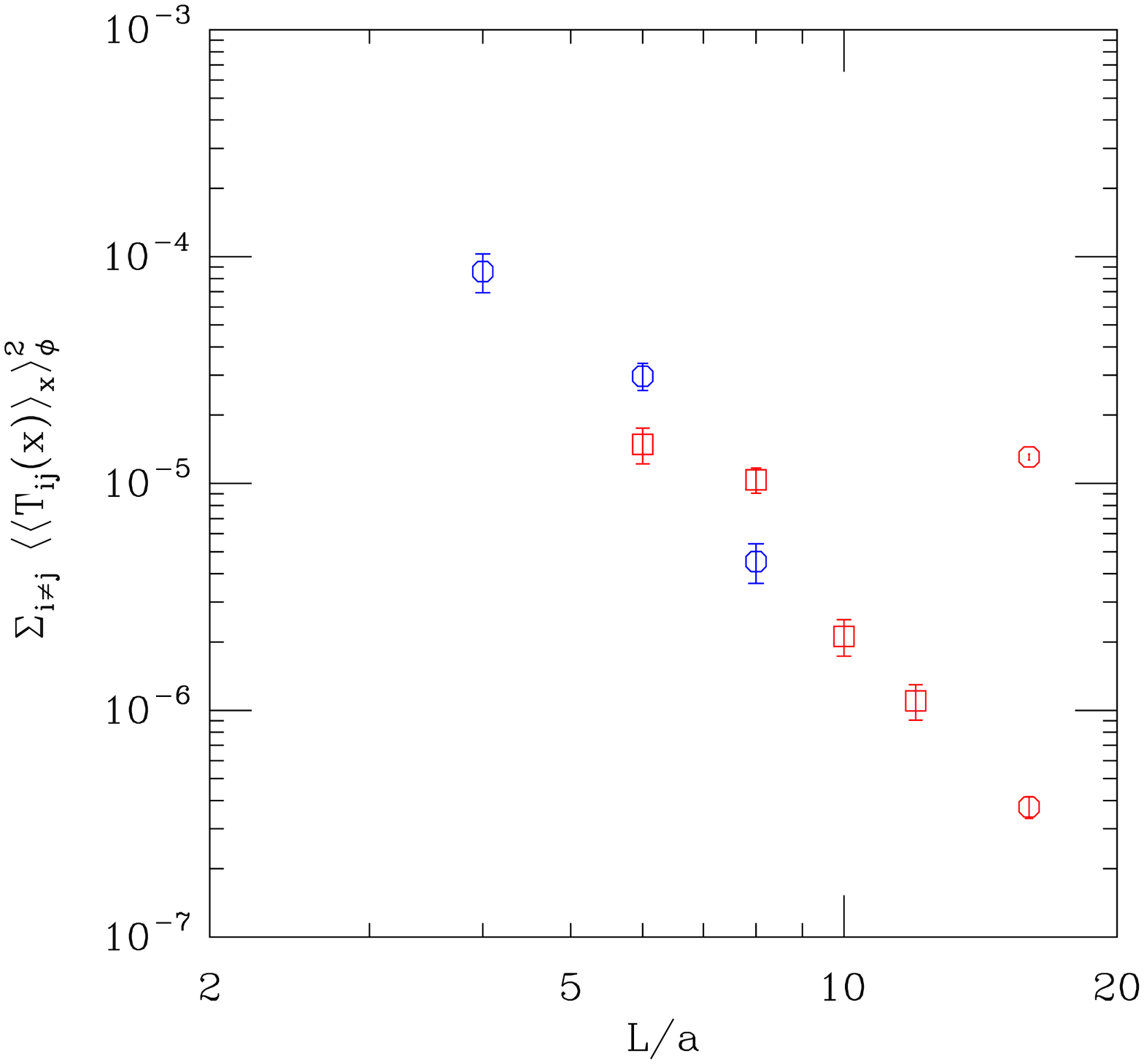}
\caption{\label{tij2} Left: zero-momentum-projected $xy$ flow-tensor 
correlators. 
Middle: Spatial correlations of $xx-zz$ stress-energy tensor. 
Right: the squared-average of $T_{ij}$ as a function of lattice extent 
$L$ (red symbols at constant physical size $L \propto 1/C$, 
circles at $C=1$; the lowest lying data fit to $\sim L^{-3.85(14)}$).}
\end{center}
\end{figure}
%%%%%%%%%%%%%%%%%%%%%%%%%%%%%%%

The left plot in Fig.~\ref{tij2} shows a zero-momentum-projected 
flow-tensor correlator (source at $t=8a$). 
While there appears to be some structure, suggesting the presence of 
a quantum ``sound mode'', the same is not seen at all lattice 
spacings (outside of the ``enhancement'' at $t=t_{source}$). 
Like with the entropy density, some components (or combinations thereof) 
of the flow and stress-energy tensors show non-zero, persistent spatial 
correlations (middle plot). 
Indeed, the off-diagonal spatial components ($T_{ij}$ and $\Omega_{ij}$) 
give non-zero averages (indicating non-zero flow) which only vanish 
in the ``thermodynamic'' limit ($L \to \infty$; see the rightmost plot). 
Even then, some ensembles appear to be rather stubborn and find a 
different minimum of the action, perhaps due to macroscopic, topological 
features (e.g., vortex rings wrapped around the torus). 
It would be interesting to search these lattices for such structures in 
order to determine to what extent such ensembles should contribute to an 
overall average (i.e., whether they are indeed due to the boundary 
conditions or not). 
That these elements are non-zero may also be a hint of existence of 
structures analogous to well-studied calorons and instantons in QCD. 
In a hydrodynamic context, these structures can be interpreted as quantum 
vortices, perhaps triggering turbulence. 
Observables such as the relativistic circulation $C_{P} = \oint_P (p+e) u_{\mu} dx^{\mu} $ across a closed 
path P can be used to investigate the existence and 
relevance of such phenomena.
 
%\begin{equation}
%\end{equation}

In conclusion, we have argued that a lattice implementation of ideal quantum 
hydrodynamics can give insight into a hitherto unexplored limit of strongly 
interacting matter, one where the dissipation vanishes but the microscopic 
and macroscopic length scales may not be well separated. 
We have discussed the technical details of this implementation and presented 
some preliminary results. 
None of the latter should be taken as anything other than a feasibility 
demonstration, this is the beginning of what can only be a very involved 
research project.

GT acknowledges the financial support received from the Helmholtz International 
Centre for FAIR within the framework of the LOEWE program 
(Landesoffensive zur Entwicklung Wissen-schaftlich-\"Okonomischer 
Exzellenz) launched by the State of Hesse. 
GT also acknowledges support from DOE under Grant No.~DE-FG02-93ER40764. 
Simulations were performed at the Institute for Theoretical Physics at the 
University of Regensburg and we thank the administrators for continued access 
and smooth machine operation. 
TB acknowledges his family's patience and support of what has become his 
late-night hobby.

\end{document}